\documentclass[12pt,preprint]{aastex}
\usepackage{epsfig}
\def\lsim{\raise0.3ex\hbox{$<$}\kern-0.75em{\lower0.65ex\hbox{$\sim$}}}
\def\gsim{\raise0.3ex\hbox{$>$}\kern-0.75em{\lower0.65ex\hbox{$\sim$}}}

\shorttitle{Star Formation in Turbulent Magnetic Clouds}
\shortauthors{Li \& Nakamura}

\begin{document}

\title{Magnetically Regulated Star Formation in Turbulent Clouds}

\author{Zhi-Yun Li\altaffilmark{1}, and Fumitaka Nakamura\altaffilmark{2}}
\altaffiltext{1}{Department of Astronomy, University of Virginia, 
Charlottesville, VA 22903; zl4h@virginia.edu}
\altaffiltext{2}{Faculty of Education and Human Sciences, Niigata University,
8050 Ikarashi-2, Niigata 950-2181, Japan; fnakamur@ed.niigata-u.ac.jp} 
\begin{abstract}

We investigate numerically the combined effects of supersonic turbulence, 
strong magnetic fields and ambipolar diffusion on cloud evolution leading 
to star formation. We find that, in clouds that are initially magnetically 
subcritical, supersonic turbulence can speed up star formation, 
through enhanced ambipolar diffusion in shocks. The speedup overcomes 
a major objection to the standard scenario of low-mass star formation 
involving ambipolar diffusion, since the diffusion time scale at the 
average density of a molecular cloud is typically longer than the 
cloud life time. At the same time, the strong magnetic field can 
prevent the large-scale supersonic turbulence from converting most of 
the cloud mass into stars in one (short) turbulence crossing time, 
and thus alleviate the high efficiency problem associated with the 
turbulence-controlled picture for low-mass star formation. We propose 
that relatively rapid but inefficient star formation results from 
supersonic collisions of somewhat subcritical gas in strongly magnetized, 
turbulent clouds. The salient features of this shock-accelerated, 
ambipolar diffusion-regulated scenario are demonstrated with numerical 
experiments. 
\end{abstract}


\section{Introduction}

The standard scenario of low-mass star formation envisions quasi-static
condensation of dense cores out of magnetically subcritical clouds 
and inside-out core collapse leading to star formation (Shu, Adams \&  
Lizano 1987; Mouschovias \& Ciolek 1999). An alternative 
is the turbulence-controlled star formation, with magnetic fields 
playing a minor role if any (Larson 1981; Mac Low \& Klessen 2004). In 
this picture, dense cores arise from compression in the supersonic 
turbulence observed in molecular clouds (Padoan et al. 2001; Gammie 
et al. 2003; Li et al. 2003). The cores so formed are transient 
entities, however. Even though some of them may have column 
density distributions resembling those of static Bonnor-Ebert spheres, 
their internal motions are typically dynamic, with transonic or even 
supersonic speeds (Ballesteros-Paredes, Klessen, \& V\'azquez-Semadeni 
2003) which are difficult to reconcile with the subsonic internal 
motions inferred in low-mass starless cores (Lee, Myers \& Tafalla 1999). 
A potentially more serious difficulty with this 
picture is the rate and efficiency of star formation. Numerical 
simulations have shown that supersonic turbulence decays in one 
free-fall time or less, with or without a strong magnetization (Mac Low 
et al. 1998; Stone, Ostriker \& Gammie 1998; Padoan \& Nordlund 1999). 
Without additional support, self-gravitating clouds would collapse in 
one free-fall time, leading to a rate of star formation well above 
that inferred on the Galactic scale (Evans 1999). Also, stars in typical 
low-mass star forming clouds contain less than a few percent of the cloud 
mass. Such a low star formation efficiency is not naturally explained. 

Excessive star formation is avoided in the standard scenario by an ordered 
magnetic field, which is postulated to provide most of the cloud support. 
The gradual weakening of magnetic support through ambipolar diffusion 
leads to core formation. The cores so formed tend to have subsonic infall 
speeds consistent with observations (Li 1999; Ciolek \& Basu 2000). 
However, calculations based on this scenario have so far avoided a direct
treatment of turbulence by starting from relatively quiescent regions of 
moderately high density (several times $10^3$~cm$^{-3}$ or higher), as 
opposed to the average cloud density (a few times $10^2$~cm$^{-3}$ or 
lower; Blitz 1993). How the over-dense regions form out of the more 
turbulent background in the first place was not addressed. It is unlikely 
for them to have condensed out quasi-statically through ambipolar diffusion, 
because the level of ionization in 
the background is enhanced over the value given by cosmic ray ionization 
alone (McKee 1989), as a result of photoionization of interstellar 
far-ultraviolet radiation field. The enhancement makes the time scale 
for quiescent ambipolar diffusion at the average density longer than the 
cloud lifetime (Myers \& Khersonsky 1995). In this Letter, we show that 
the supersonic turbulence observed in molecular clouds can speed up 
ambipolar diffusion in localized regions through shock compression
without turning most of the cloud mass into stars in one crossing time. 

\section{Problem Formulation}
\label{sec:method}

We consider strongly magnetized, sheet-like clouds, taking advantage of 
the tendency for cloud material to settle along field lines into a 
flattened configuration. The basic formulation we adopt here is the 
same as that of Nakamura \& Li (2002, 2003) and Li \& Nakamura (2002), 
which was originally developed for treating ambipolar diffusion-driven 
fragmentation of quiescent magnetic clouds in the presence of small 
perturbations (see also Indebetouw \& Zweibel 2000 and Basu \& Ciolek
2004). Briefly, we assume that the fluid motions are confined 
to the plane of the sheet, with force balance maintained in the 
vertical direction at all times (Fiedler \& Mouschovias 1993). The 
cloud evolution is governed by a set of vertically integrated MHD 
equations that include ambipolar diffusion. We consider the case where 
the magnetic field is coupled to the bulk neutral cloud material 
through ion-neutral collision, with ion density proportional to the 
square root of neutral density. A canonical value of $3\times 
10^{-16}$cm$^{-3/2}$g$^{1/2}$ is adopted for the proportionality 
constant (Elmegreen 1979). We solve 
the hydrodynamic part of the governing equations using Roe's TVD 
method, and determine the gravitational and magnetic potentials 
outside the sheet using a convolution method based on FFT. The 
computation timestep is chosen small enough to satisfy both the 
CFL condition and the stability condition for treating ambipolar 
diffusion explicitly. 

We consider an initially uniform mass distribution in the $x$-$y$ plane, 
with column density $\Sigma_0=4.68\times 10^{-3} A_V$~g~cm$^{-2}$ (where 
$A_V$ is visual extinction for standard grain properties). The cloud is 
assumed isothermal, with sound speed $c_s=1.88\times 10^4 T_{10}^{1/2}
$~cm~s$^{-1}$ (where $T_{10}$ is the temperature in units of 10~K). 
Above a critical column density $\Sigma_c=10^2 \Sigma_0$, the equation 
of state is stiffened to enable simulation beyond the 
formation of the first ``collapsed'' object. The Jeans length 
$L_J=c_s^2/(G\Sigma_0)=0.37\ T_{10}/A_V$~pc and gravitational collapse 
time $t_g=L_J/c_s=1.90\times 
10^6 T_{10}^{1/2}/A_V$~yr provide two basic scales for our problem. 
We adopt a computation box of size $10\; L_J$ (corresponding to a cloud 
mass $10^2$ times the Jeans mass $M_J\equiv \Sigma_0 L_J^2 =3.02\ T_{10}
^2/A_V$~$M_\odot$), and impose periodic conditions at the boundaries. 
The cloud is threaded vertically by a uniform magnetic field, 
characterized by the flux-to-mass ratio $\Gamma_0$ in units of the 
critical value $2\pi G^{1/2}$. Only strongly magnetized clouds with 
$\Gamma_0$ close to or exceeding unity are considered, in keeping with 
the adopted thin-sheet approximation.   

To mimic the turbulent motions observed in molecular clouds, we stir 
the cloud with a supersonic velocity field at $t=0$. Following the 
usual practice (e.g., Stone et al. 2000), we prescribe the turbulent 
velocity field in Fourier space, with a power spectrum $v_k^2 \propto 
k^{-4}$. Other choices yield 
qualitatively similar results, as long as the power is dominated by 
large-scale motions. Numerically, we adopt a $256\times 256$ grid, 
and consider only discrete values $k_i=2\pi i/L$ (where $L$ is the 
box size and $i=1$, 2, ..., N-1, with $N=256$) for wavenumber 
components, $k_x$ and $k_y$. For each pair of $k_x$ and $k_y$, we 
randomly select an amplitude from a Gaussian 
distribution consistent with the power spectrum for the total 
wavenumber $k=\sqrt{k_x^2+k_y^2}$ and a phase between 0 and $2\pi$. 
The resulting field is then transformed into the physical space.  
The spatial distributions of $V_x$ and $V_y$ are generated 
independently, and the final velocity field is scaled such that the 
rms Mach number of the flow ${\cal M}$ has a prescribed value. The 
velocity fields so specified tend to be strongly 
compressive. Adopting an incompressible turbulent velocity field 
does not change the result qualitatively. 

\section{Results} 

Two representative clouds will be considered in detail, one magnetically
subcritical with $\Gamma_0=1.2$ and the other supercritical with $\Gamma
_0=0.8$. We choose a Mach number ${\cal M}=10$, which yields a turbulence 
crossing time $t_{\rm x}\equiv 10 L_J/(2{\cal M} c_s)=0.5\; t_g$. The 
results are displayed in Figs.~1 and 2. Fig.~1 presents the snapshots of 
the clouds at various times, and 
Fig.~2 shows,  as a function of time, the mass fraction of the regions 
that have become more than 10 times denser (in column density) than 
the average {\it and} are magnetically supercritical. These so-called 
``dense supercritical'' 
regions contain both ``collapsed'' objects whose contraction has been 
arrested by the stiffening of the equation of state 
and isothermal core material well on its way to star 
formation. We will use their mass fraction as a measure of the efficiency 
of star formation, bearing in mind however that probably only a fraction 
(say 1/3 to 1/2) of the dense supercritical material would be turned into 
stars in the presence of powerful protostellar outflows (Matzner \& McKee 
2000; Shu, Li \& Allen 2004). 

We begin with the subcritical cloud. Its snapshots are
displayed in the first five panels of Fig.~1. Panel (a) shows the cloud 
after about one turbulence crossing time (at $t=0.448$, in units of 
$t_g$ here and below), when a large 
fraction of the cloud mass has been compressed into a network of 
filaments. The filaments would collapse quickly were it not for the 
strong magnetic fields trapped in them. By $t=0.955$  
(panel b), they are well on their way to re-expansion, 
leaving behind little dense supercritical material (which has a mass 
faction of order $5\%$ according to Fig.~2). Although most of the shocked 
regions do not collapse right away, they are permanently altered. Their 
flux-to-mass ratios are changed by ambipolar diffusion, whose rate is 
enhanced by the increase in both density (which weakens the field-matter 
coupling) and gradient of the field strength. The magnetically altered 
regions cannot 
return to their pre-compression, uniformly magnetized state after 
shock passage. Rather, a magnetically differentiated structure 
develops, with regions of relatively low flux-to-mass ratio embedded 
within a more strongly magnetized background. The supercritical 
``islands'' are marked by white contours in each panel. These regions 
have a head-start over their subcritical surroundings in forming 
stars, through further turbulent compression and/or ambipolar diffusion. 
(We have tested our standard simulation of $\Gamma_0=1.2$ in the 
limit of zero ambipolar diffusion, and found that the $B/\Sigma$ 
ratio remains constant to within $10^{-4}$.)

After the initial phase of strong compression and re-expansion, the 
cloud settles down to a more relaxed configuration. An example of
this new phase of cloud evolution is shown in panel~(c). Prominent 
at this time ($t=1.59$ or $3.0$~Myrs, adopting here and below 
the fiducial combination
$T_{10}=1$ and $A_V=1$, appropriate for Taurus clouds according to
Arce \& Goodman 1999) are several supercritical 
filaments, the densest parts of which have already begun dynamic 
contraction and formed ``collapsed'' objects. Note that the dense
star-forming regions are well separated, reminiscent of the 
dispersed mode of star formation observed in Taurus-like clouds. 
Once created by large-scale compression at well separated 
locations, the dense regions stay apart because their  
mutual gravitational attraction is cancelled to a large extent
by the magnetic repulsion between them. This mode is 
illustrated further in panel~(d), where $t=2.88$ (or $5.5$~Myrs) and 
$\sim 10\%$ of the cloud mass now resides in the dense supercritical
regions. Note that the dense ``ridge'' at the lower-left corner has 
a much lower velocity than its surrounding medium. This is also true 
for other high density regions in general, broadly consistent with 
the observation that low-mass star forming cores are more quiescent 
than their surroundings. 

The mass fraction of dense supercritical regions increases steadily, 
as a result of continued ambipolar diffusion. The steady increase 
is punctuated by bumps at late times (see Fig.~2). These bumps are 
produced by interactions of dense regions, which are evident in 
panel~(e) of Fig.~1, where $t=9.55$ (or 18~Myrs). The interactions 
become more important as 
more cloud material becomes supercritical, since the gravitational 
attraction between supercritical regions is not completely offset 
by magnetic repulsion. The unbalanced ``effective'' gravity tends 
to drive the supercritical regions together, which enhances the 
possibility for further interactions. The clustering of ``heavier'' 
supercritical regions against a more strongly magnetized background 
may, given enough time, drive the cloud from the initial dispersed 
mode of star formation to a clustered mode of star formation. In the 
regions where the dense supercritical material preferentially gathers, 
the local efficiency of star formation can be much higher than the
average. 

The magnetically supercritical cloud evolves differently. Its snapshots
are displayed in the last three panels of Fig.~1. Panel~(f) shows the 
cloud at $t=0.487$, when a network of filaments has formed, as in 
the subcritical case, although the filaments here are thinner, because 
of a weaker magnetic resistance to shock compression. These filaments 
are sufficiently supercritical that, when enough mass is accumulated, 
they break up gravitationally into a string of dense 
cores along their length, unlike their more magnetized counterparts. 
By the time shown in panel~(g) ($t=0.955$), some of the dense cores 
have already merged together, creating several dense supercritical 
regions. These regions contain about half of the cloud mass (see 
Fig.~2). They dominate the subsequent cloud evolution through 
mutual gravitational interactions. By the time shown in panel~(h) 
($t=3.50$), some of the dense regions have merged together, and 
new blobs have formed through interactions. Most of the dense regions  
are clustered together gravitationally. The clustering is retarded 
only moderately by magnetic repulsion, since the cloud is magnetically 
supercritical as a whole, and the dense regions are even more so. 
Nearly $20\%$ of the cloud mass does become  subcritical, however, as 
a result of magnetic flux diffusing from the high-density to low-density 
regions. It may explain why the mass fraction of the dense supercritical 
regions hovers around half, rather than a value closer to unity as one 
might naively expect for supercritical clouds. 

We have explored the effects of the same (strong) turbulence on clouds
with other degrees of magnetization, including one that is critically
magnetized everywhere initially ($\Gamma_0=1$). Dense supercritical 
regions are created promptly in this cloud, with a mass fraction 
intermediate between that of the subcritical ($\Gamma_0=1.2$) and 
supercritical ($\Gamma_0=0.8$) cloud, as shown in Fig.~2. Also shown 
in Fig.~2 is a more subcritical cloud with $\Gamma_0=1.5$. It does 
not produce a dense supercritical object until $t\sim 5$ (or $\sim 
10$~Myrs). Such a long dormant time would argue against star-forming 
clouds being subcritical by a large factor everywhere, although a 
stronger turbulence can in principle induce localized cloud collapse and 
star formation sooner. In practice, a distribution of the flux-to-mass 
ratio $\Gamma_0$ is expected in realistic clouds. The least magnetized 
regions are expected to collapse first, either on their own (if they 
are supercritical and self-gravitating) or through turbulent compression. 
In such a case, the star formation rate and efficiency would obviously 
depend on the $\Gamma_0$ distribution, which is unknown at present. 

\section{Discussion and Conclusions}

We have demonstrated that star formation with a relatively low efficiency 
can occur over an extended period of time in highly turbulent, somewhat 
magnetically subcritical clouds. The low efficiency is made possible by 
the strong magnetic field, which prevents the global cloud collapse in 
one turbulence crossing time. The supersonic turbulence, on the other 
hand, speeds up ambipolar diffusion in localized regions through shock 
compression. The acceleration of ambipolar diffusion by turbulence has 
been examined analytically by Kim \& Diamond (2002), Zweibel (2002) and 
Fatuzzo \& Adams (2002), and numerically by Heitsch et al. (2004). Our 
investigation differs from these studies in adopting a highly compressive
turbulent flow and in that the cloud dynamics are computed rather than
prescribed. 

To further illustrate the role of turbulence in star formation, we have 
rerun the subcritical ($\Gamma_0=1.2$) case, varying the turbulent Mach 
number ${\cal M}$ from 0.3 to 10. The result is plotted in Fig.~3. For 
${\cal M}\lsim 1$, it takes an order of magnitude longer than the 
gravitational collapse time to produce the first dense magnetically 
supercritical object. As ${\cal M}$ increases, dense supercritical 
regions appear earlier. Beyond some critical value ($M_c\approx 5$ for 
our particular examples), such regions form promptly, within one collapse 
time. The ability for a 
subcritical cloud to form stars quickly in the presence of 
a supersonic turbulence removes a major objection to the standard scenario 
for low-mass star formation through ambipolar diffusion based on cloud 
lifetime arguments (Hartmann, Ballesteros-Paredes, \& Bergin 2001). 

Our calculations support the idea of bimodal star formation (Shu et al. 
1987). In magnetically subcritical clouds, dense supercritical regions 
are created by large-scale turbulence at well separated locations. They 
stay separated until a large enough fraction of the cloud mass becomes 
supercritical for the gravitational attraction to overwhelm the magnetic 
repulsion. We identify this mode of star formation with the mode of 
relatively inefficient, dispersed star formation exemplified by the 
Taurus clouds. If the cloud lives long enough, a large fraction of the 
cloud mass will become supercritical sooner or later, through continued 
ambipolar diffusion. In regions where the supercritical materials 
collect gravitationally, the star formation efficiency could be much 
higher than the general background. Such regions could be the sites 
of efficient cluster formation. Efficient star formation occurs much
more quickly if the cloud is initially magnetically supercritical, 
as in our second example. Our simulations suggest that a realistic 
turbulence with power dominated by large-scale motions does not 
fundamentally change the bimodal nature of star formation in magnetic 
clouds. 3D calculations are needed to firm up this conclusion.

\acknowledgments 
We thank C. Gammie and E. Ostriker for helpful discussion. Support for 
this work was provided by Grant-in-Aid for Scientific Research 
(No. 15740117) by the Ministry of Education, Culture, Science and 
Technology of Japan and NSF grant AST-0307368.

\clearpage

\begin{figure}
\epsscale{1.0} 

\plotone{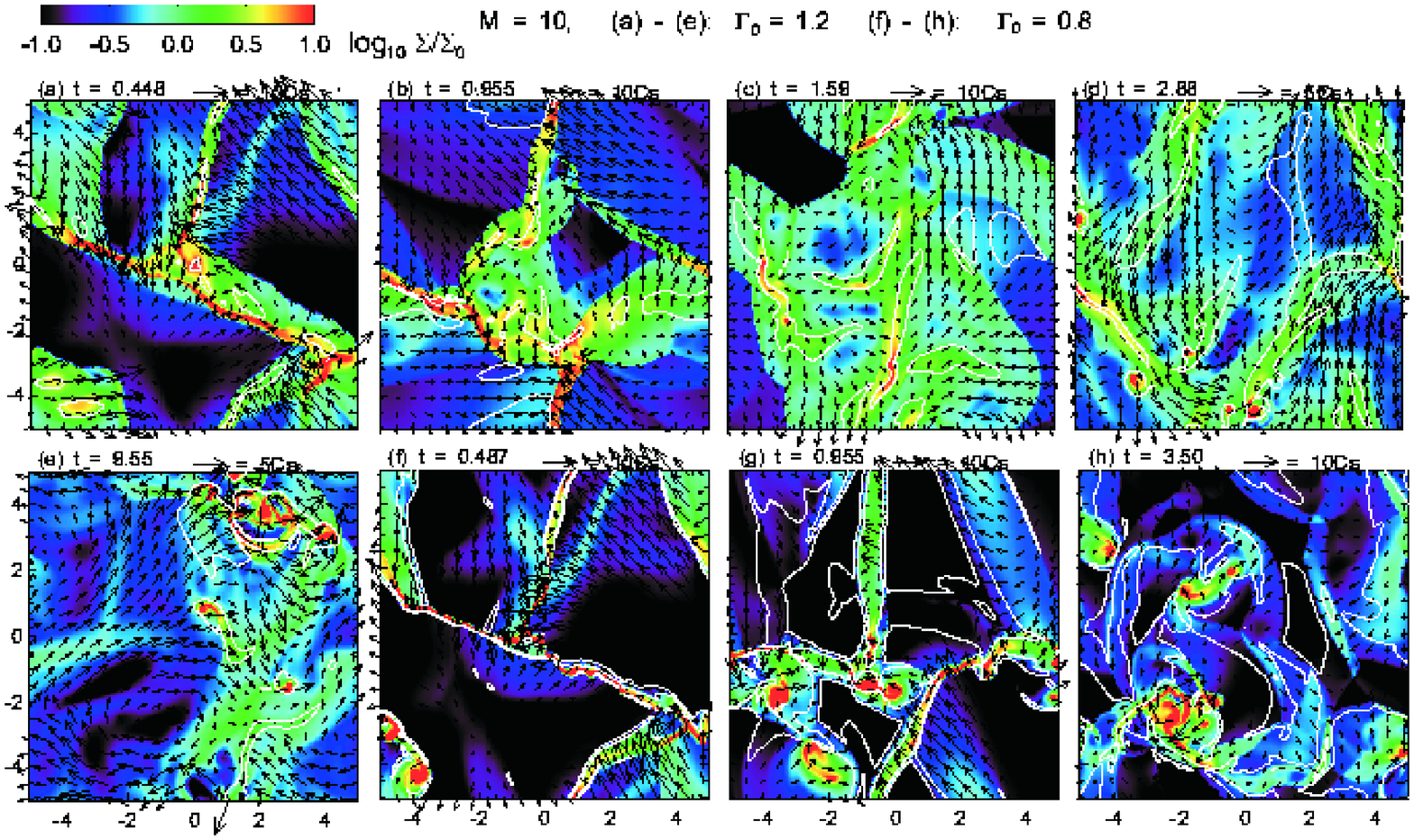}

\caption{Snapshots of magnetized turbulent clouds. Panels (a) through (e) 
are for an initially magnetically subcritical 
cloud with $\Gamma_0=1.2$, and (f) through (h) for a supercritical cloud 
with $\Gamma_0=0.8$. The clouds are stirred at $t=0$ by the same random 
velocity field of Mach number ${\cal M}=10$. The color bar is for column 
density (in units of the initial value $\Sigma_0$). The time labeled 
is in units of the gravitational collapse time $t_g$, and the length unit 
is the Jeans length $L_J$. Shown in each panel are 
contours of critical flux-to-mass ratio (in white) and the velocity field 
(in arrows). The arrow length is proportional to the flow speed, with 
normalization indicated above the panel. 
\label{fig:1}}
\end{figure}

\begin{figure}
\epsscale{0.6}
\plotone{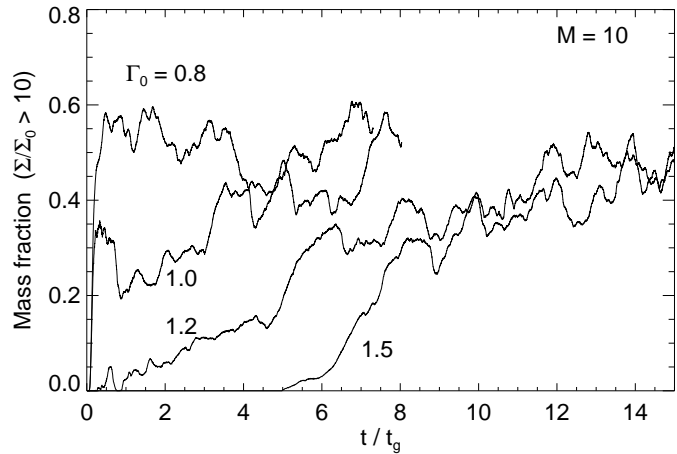}
\caption{Time evolution of the mass fraction of the regions more
than 10 times denser (in column density) than the average that 
are also magnetically supercritical, showing the effects of the 
degree of cloud magnetization $\Gamma_0$ (labeled beside each 
curve). All cases have the same initial random velocity field of 
Mach number ${\cal M}=10$.   
\label{fig:2}}
\end{figure}

\begin{figure}
\epsscale{0.6}
\plotone{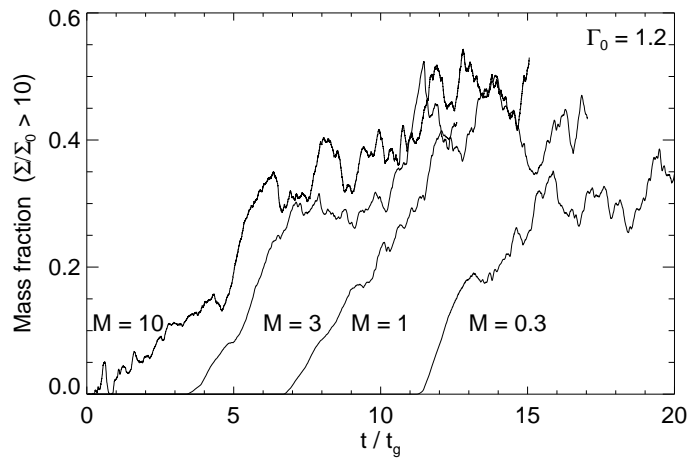}
\caption{Time evolution of the mass fraction of the dense regions that 
have become magnetically supercritical through ambipolar diffusion in 
a subcritical cloud ($\Gamma_0=1.2$), showing the effects of the 
level of turbulence ${\cal M}$. }
\end{figure}

\end{document}